%% file: main.tex
\documentclass{ifacconf}

\usepackage{graphicx}      
\usepackage{natbib}        
\usepackage{amsmath, amssymb, gensymb, bm, mathtools}
\usepackage{xcolor}
\input{macros}
\begin{document}
\begin{frontmatter}

\title{A Minimum Energy Filter for Distributed Multirobot Localisation\thanksref{footnoteinfo}} 

\thanks[footnoteinfo]{This research is supported by the Commonwealth of Australia as represented by the Defence Science and Technology Group of the Department of Defence and by the Australian Research Council Discovery Project DP190103615: ``Control of Network Systems with Signed Dynamical Interconnections''}

\author[First]{Jack Henderson} 
\author[First]{Jochen Trumpf} 
\author[Second]{Mohammad Zamani}

\address[First]{Research School of Electrical, Energy and Materials Engineering, Australian National University\\E-mail: \{jack.henderson, jochen.trumpf\}@anu.edu.au}
\address[Second]{Land Division, Defence Science and Technology Group, Australia\\E-mail: mohammad.zamani@dst.defence.gov.au}

\input{sections/0_abstract}

\end{frontmatter}

\input{sections/1_introduction}
\input{sections/2_background}
\input{sections/3_problem_formulation}
\input{sections/4_results}

\input{sections/5_simulations}
\input{sections/6_conclusions}



\bibliography{references}             
\appendix
\end{document}

%% file: macros.tex
\newcommand{\SE}{\mathrm{SE}}
\newcommand{\se}{\mathfrak{se}}
\newcommand{\SO}{\mathrm{SO}}
\newcommand{\so}{\mathfrak{so}}

\newcommand{\inv}{^{\text{-}1}}

\newcommand*{\Real}[1]{\ensuremath{\mathbb{R}^{#1}}}

\newcommand\norm[1]{\ensuremath{\left\Vert #1 \right\Vert}}
\newcommand\proj[1]{\ensuremath{\mathbb{P}\left( #1 \right)}}
\newcommand\projsym[1]{\ensuremath{\mathbb{P}_s\left( #1 \right)}}
\newcommand\projasym[1]{\ensuremath{\mathbb{P}_a\left( #1 \right)}}

\DeclareMathOperator*{\argmin}{arg\,min}

\DeclareMathOperator{\blkdiag}{blkdiag}

\DeclareMathOperator{\rank}{rank}
\DeclareMathOperator{\trace}{tr}
\DeclareMathOperator{\vex}{vex}
\newcommand\Hess{\mathrm{Hess}}

\newcommand{\doubleplus}{\mathrel{+\!\!\!+}}

%% file: sections/0_abstract.tex
\begin{abstract}                
We present a new approach to the cooperative localisation problem by applying the theory of minimum energy filtering. We consider the problem of estimating the pose of a group of mobile robots in an environment where robots can perceive fixed landmarks and neighbouring robots as well as share information with others over a communication channel. Whereas the vast majority of the existing literature applies some variant of a Kalman Filter, we derive a set of filter equations for the global state estimate based on the principle of minimum energy filtering. We show how the filter equations can be decoupled and the calculations distributed among the robots in the network without requiring a central processing node. Finally, we provide a demonstration of the filter's performance in simulation.
\end{abstract}

%% file: sections/1_introduction.tex
\section{Introduction}
In a wide range of robotics applications, an accurate estimation of the current position and orientation (pose) of a robot is essential for its primary mission. However, in environments where sensor performance is degraded, traditional approaches to state estimation begin to fail, for example when GNSS systems are actively jammed or robots have very limited access to external landmarks.

Collaborative localisation (CL) is an approach that can be utilised where multiple robots are operating in the same environment. The core concept is that the information gained by sensors on one robot can be shared with other robots in the environment thereby increasing the accuracy of the pose estimates. This is useful in the aforementioned case of sensor degradation, and can also be used for groups of heterogeneous robots where robots have different sensing capabilities. 

When information such as local state estimates are shared between robots in a filtering framework, the pose estimates of each robot are no longer independent. If this state-dependency is not properly accounted for, it can lead to data incest and over-confidence problems \citep{howardPuttingTeamEgocentric2003}. 

One solution to this problem is presented by \cite{roumeliotisDistributedMultirobotLocalization2002}, who derive a centralised Extended Kalman Filter (EKF) which jointly estimates the pose of all robots in the network and where all of the information dependencies are tracked in a single covariance matrix. They go on to show that the joint filter equations can be decoupled into a set of smaller, communicating filters distributed among the robots in the network.

There are two drawbacks to this type of approach. Firstly, the entire joint covariance matrix must be tracked, either in a centralised or distributed arrangement. Secondly, after each measurement step, a robot must communicate to all other robots in the network to update the estimates and covariance terms that are tracked by other robots. This is often impractical in most scenarios as a robust, fully-connected communications network topology cannot be guaranteed.

Recent approaches to CL have focused on reducing the data that is tracked by each robot and relaxing the communication network topology constraints. For example, \cite{carrillo-arceDecentralizedMultirobotCooperative2013} proposes a filter local to each robot which only tracks the local state. Cross-covariance terms are not tracked and are instead estimated locally when two robots meet and share information. 
While this reduces the communication overhead, it comes at the cost of being too conservative in the dependency estimation and not utilising available data to the maximal possible extent. Their experimental results show the performance of this type of filter is worse than the joint EKF.
Further work by \cite{luftRecursiveDecentralizedLocalization2018} aims to better approximate the cross covariance terms, and demonstrates performance very similar to, although slightly worse than, the joint EKF from \cite{roumeliotisDistributedMultirobotLocalization2002}.

The common element in these works is that they all utilise the Extended Kalman Filter and are all compared to the joint EKF presented by \cite{roumeliotisDistributedMultirobotLocalization2002} as it provides a baseline estimate given no restrictions on computation or communication.
While the EKF and variants such as the Multiplicative EKF (MEKF) \citep{markleyAttitudeErrorRepresentations2003} are industry standard in terms of pose filtering algorithms, there are potentially better alternatives. All variants of the EKF rely on linearising the system, which can result in instability and convergence issues when estimating highly non-linear systems such as robot pose kinematics. More recently, the approach of minimum energy filtering has been demonstrated by \cite{zamaniMinimumEnergyFilteringAttitude2013} as a more accurate and robust algorithm for pose estimation. A collaborative minimum-energy pose estimation algorithm has been proposed by \cite{zamaniCollaborativePoseFiltering2019}, but opts for estimating cross-covariance terms rather than tracking the full state of the system.

In this paper, we present the derivation of a centralised geometric approximate minimum-energy (GAME) filter to estimate the poses of a network of robots using both interoceptive and exteroceptive measurements. We show how this filter can be equivalently derived as a set of collaborative filters which run locally on each robot in the network. The decoupled filters provide exactly the same pose estimates as the centralised filter and information-sharing is only necessary during the filter update step, where exteroceptive measurements are processed.

The filter we derive provides a baseline which can be used as a benchmark for future implementations of other minimum-energy filters, much in the same way that \cite{roumeliotisDistributedMultirobotLocalization2002} has been used as the baseline for further developments of EKF-based filters.

The remainder of this paper is organised as follows. In Section \ref{sec:preliminaries} we briefly introduce a number of concepts and the notation used in the paper. Section \ref{sec:problem_formulation} formally states the problem we aim to solve and we present our solution in Section \ref{sec:results}. We then demonstrate an implementation of the filter through a simulation in Section \ref{sec:simulations} and conclude the paper in Section \ref{sec:conclusion}.

%% file: sections/2_background.tex
\section{Preliminaries}
\label{sec:preliminaries}
In this section we introduce the notation and conventions used throughout the paper.

\subsection{Notation}
$(.)^\top$ denotes the matrix transpose, $\bm{I}_n$ denotes an $n\times n$ identity matrix. The operators $\exp$ and $\log$ denote the matrix exponential and matrix logarithm respectively.

We use the matrix Lie groups $\SO(3)$ to represent rotations and $\SE(3)$ to represent poses in homogeneous coordinates. The corresponding Lie algebras are $\so(3)$ and $\se(3)$, respectively.
\begin{align}
\SO(3) &= \left\{R \in \Real{3\times 3}~\vert~R^\top R = \bm{I}_3,~\det{R} = 1 \right\}
\\
\SE(3) &= \left\{ X =  \begin{bmatrix}R & p\\ 0_{1 \times 3} & 1\end{bmatrix} ~ \vert ~ R \in \SO(3), ~ p \in \Real{3}  \right\}
\\
\so(3) &= \left\{ \Psi \in \Real{3\times 3} ~ \vert ~ \Psi^\top = -\Psi  \right\}
\\
\se(3) &= \left\{  \Gamma = \begin{bmatrix} \Psi & v \\ 0_{1\times 3} & 0 \end{bmatrix} ~ \vert ~ \Psi \in \so(3), ~ v \in \Real{3} \right\}
\end{align}

We define the following maps which allow us to switch between matrix and vector representations.
\begin{align}
(.)_\times &: \Real{3} \rightarrow \so(3)
&
\omega_\times &:= \begin{bmatrix} 0 & -\omega_3 & \omega_2 \\ \omega_3 & 0 & -\omega_1 \\ -\omega_2 & \omega_1 & 0 \end{bmatrix}
\\
\vex &: \so(3) \rightarrow \Real{3}
&
\vex(\Omega) &:= \vex(\omega_\times) = \omega
\\
(.)^\wedge &: \Real{6} \rightarrow \se(3)
&
\gamma^\wedge &:= \begin{bmatrix}
\omega_\times & v \\ 0 & 0
\end{bmatrix} = \Gamma
\\
(.)^\vee &: \se(3) \rightarrow \Real{6}
&
\Gamma^\vee &:= \left(\gamma^\wedge\right)^\vee = \gamma
\\
(.)^\curlyvee &: \se(3)^n \rightarrow \Real{6n}
&
\bm{\Gamma}^\curlyvee &= \left[ \Gamma_1^{\vee\top}, \ldots, \Gamma_n^{\vee\top}  \right]^\top
\\
(.)^\curlywedge &: \Real{6n} \rightarrow \se(3)^n
&
\bm{\gamma}^\curlywedge &:= \left(\bm{\Gamma}^\curlyvee\right)^\curlywedge = \bm{\Gamma}
\end{align}
where
\begin{align}
\gamma &= \begin{bmatrix}
\omega \\ v
\end{bmatrix}, ~ \omega, v \in \Real{3}
&
\Gamma &\in \se(3)
\\
\bm{\gamma} &\in \Real{6n}
&
\bm{\Gamma} &= \left(\Gamma_1, \ldots, \Gamma_n \right) \in \se(3)^n
\end{align}

The following maps are useful when working in homogeneous coordinates.
\begin{align}
\bar{(.)} &: \Real{3} \rightarrow \Real{4}
&
\bar{v} &:= \begin{bmatrix} v^\top & 1 \end{bmatrix}^\top
\\
\mathring{(.)} &: \Real{3} \rightarrow \Real{4}
&
\mathring{v} &:= \begin{bmatrix} v^\top & 0 \end{bmatrix}^\top
\\
\tilde{(.)} &: \Real{3\times 3} \rightarrow \Real{4 \times 4}
&
\tilde{M} &:= \begin{bmatrix}M & 0 \\ 0 & 1 \end{bmatrix}
\end{align}

Observe the following identities for $\Gamma \in \se(3), v \in \Real{3}$.
\begin{align}
\Gamma \bar{v} &= F(v) \Gamma^\vee & F(v) &:= \begin{bmatrix} -v_\times & \bm{I}_3 \\ 0 & 0 \end{bmatrix}
\label{eq:def_F}
\\
\Gamma^\top \bar{v} &= G(v) \Gamma^\vee & G(v) &:= \begin{bmatrix} v_\times & 0 \\ 0 & v^\top \end{bmatrix}
\label{eq:def_G}
\end{align}
We define the symmetric and skew-symmetric projections of $\Real{n\times n}$, $\mathbb{P}_s$ and $\mathbb{P}_a$ respectively, and the unique orthogonal projection, $\mathbb{P}$, of $\Real{4 \times 4}$ onto $\se(3)$ with respect to the Frobenius inner product.
\begin{align}
\mathbb{P}_s &: \Real{n \times n} \rightarrow \text{sym}(n)
&
\projsym{M} &:= \frac{1}{2}(M + M^\top)
\\
\mathbb{P}_a &: \Real{n \times n} \rightarrow \so(n)
&
\mathbb{P}_a(M) &:= \frac{1}{2}(M - M^\top)
\end{align}
\begin{gather}
\mathbb{P} : \Real{4 \times 4} \rightarrow \se(3)\quad
\proj{\begin{bmatrix}A_{3\times 3} & B_{3 \times 1} \\ C_{1\times 3} & D_{1\times 1} \end{bmatrix}} := \begin{bmatrix} \projasym{A} & B \\ 0 & 0 \end{bmatrix}
\end{gather}
Lastly, we define the element-wise multiplication operator, $\odot$, for a general group, $G$.
\begin{gather}
\odot : G^n \times G^n \rightarrow G^n
\\
(x_1, \ldots, x_n) \odot (y_1, \ldots, y_n) := (x_1 y_1, \ldots, x_n y_n)
\end{gather}
We will omit the $\odot$ symbol when the meaning is clear from context.

\subsection{Metrics}

Let $T_X\SE(3)$ denote the tangent space to the manifold $\SE(3)$ at the point $X$. Note that the Lie algebra $\se(3)$ coincides with $T_{\bm{I}}\SE(3)$ and that for all $\Gamma \in \se(3)$, the tangent vector $X\Gamma \in T_X\SE(3)$.

Let the metric $\langle ., . \rangle_X : T_X\SE(3) \times T_X\SE(3) \rightarrow \Real{}$ denote the standard left-invariant Riemannian metric on $\SE(3)$, that is
\begin{align}
\langle X\Gamma, X\Psi \rangle_X &= \langle \Gamma, \Psi \rangle_I = \langle \Gamma, \Psi \rangle
\\
&= \trace \left( \begin{bmatrix}0.5\bm{I}_3 & 0 \\ 0 & 1\end{bmatrix} \Gamma^\top \Psi \right)
\label{eq:def_trace_norm}
\\
&= \langle \Gamma^\vee, \Psi ^\vee \rangle = \left(\Gamma^\vee\right)^\top\Psi^\vee
\end{align}
 for $\Gamma, \Psi \in \se(3)$.
 
We define the distance function, $d_P$, applied to two elements $\bm{X}_1, \bm{X}_2 \in \SE(3)^n$, weighted by a positive definite matrix $P \in \Real{6n\times 6n} \succ 0$ as
\begin{align}
d_P(\bm{X}_1, \bm{X}_2) &:= \sqrt{\langle P \log(\bm{X}_2\inv \odot \bm{X}_1 )^\curlyvee, \log(\bm{X}_2\inv \odot \bm{X}_1) ^\curlyvee \rangle}.
\end{align}

\subsection{Differential Geometric Notation}
Let $f : \SE(3) \rightarrow \Real{}$ denote a differentiable map. Then $\mathcal{D}_X f(X): T_X\SE(3) \rightarrow \Real{}$ denotes the Fr\'{e}chet derivative and we have
\begin{gather}
\mathcal{D}_X f(X) \circ (X\Gamma) = \langle \nabla_X f(X), X\Gamma \rangle_X
\label{eq:def_derivative}
\end{gather}
where $X \Gamma \in T_X\SE(3)$ denotes the tangent direction in which the derivative is evaluated and $\nabla_X f(X)$ denotes the gradient at the point X with respect to the metric $\langle .,. \rangle_X$.

The second order differential map $\mathcal{D}_X^2 f(x) : T_X\SE(3) \times T_X\SE(3) \rightarrow \Real{}$ is defined as
\begin{align}
\mathcal{D}_X^2 f(x) \circ (X\Gamma, X\Psi) &= \langle \Hess_X f(X) \circ (X\Psi), X\Gamma \rangle_X
\\
&= \langle \Hess_X f(X) \circ (X\Gamma), X\Psi \rangle_X
\end{align}
where $\Hess_X f(X)$ denotes the Hessian operator.
The map can also be written in terms of first-order derivatives:
\begin{gather}
\begin{split}
\mathcal{D}_X^2 f(x) \circ (X\Gamma, X\Psi) = \mathcal{D}_X ( \mathcal{D}_X f(X) \circ (X\Gamma)) \circ (X\Psi)
\\- \langle \nabla_X f(X), X \bm{\Lambda}_\Psi(\Gamma)\rangle_X
\end{split}
\end{gather}
where $\bm{\Lambda}_\Psi: \se(3) \rightarrow \se(3)$ is the connection function. In this paper, we use the symmetric Cartan connection:
\begin{gather}
\bm{\Lambda}_\Psi(\Gamma) 
:= \frac{1}{2}(\Psi\Gamma - \Gamma \Psi)
\end{gather}

%% file: sections/3_problem_formulation.tex
\section{Problem Formulation}
\label{sec:problem_formulation}
We consider $n$ mobile robots in a fully-connected network with node set $N := \{1, \ldots, n\}$.
A set of $r$ landmarks $L := \{l_i \in \Real{3} ~\vert ~i = 1, \ldots, r \}$ are placed in the environment at fixed locations.
Each robot is equipped with a suite of interoceptive and exteroceptive sensors as well as a method to communicate directly to other robots in the network.
We aim to derive a deterministic second-order approximate minimum energy filter to estimate the pose of each robot in the network.
Initially, this will be formulated as a set of centralised equations but we will show how the filter can be decoupled and distributed among the robots in the network.

\subsection{Kinematics}
The rotation, $R_i$, and translation, $p_i$, of each robot $i \in N$ with respect to a fixed reference frame is represented as a $4\times4$ homogeneous matrix, $X_i$. The pose has the following left-invariant kinematics.
\begin{align}
X_i &= \begin{bmatrix}R_i & p_i\\ 0 & 1\end{bmatrix} \in \SE(3) & \Omega_i &= \begin{bmatrix} \omega_i \\ v_i \end{bmatrix} ^\wedge \in \se(3)\\
\dot{X}_i &= X_i \Omega_i & X_i(0) &= X_{i,0}
\end{align}
where $\omega_i$ and $v_i$ are the angular and linear velocities of the robot with respect to the reference frame.
\subsection{Measurements}
\label{sec:measurements}
A robot, $i$, can independently measure its own velocity. The measurement, $u_i$, is corrupted by zero-mean sensor noise, $\epsilon_i \in \Real{6}$.
\begin{align}
u_i &= \begin{bmatrix}
\omega_i \\
v_i
\end{bmatrix} + B_i\epsilon_i
\label{eq:velocity_measurement_def}
\end{align}
where $ B_i \in \Real{6 \times 6}$ is determined by the sensor properties.

Each robot is equipped with a sensor that measures the relative translation between the robot and landmarks in the environment. A measurement, $y \in \Real{3}$, of the landmark $l\in L$ taken by robot $i$ is corrupted by zero-mean sensor noise, $\delta \in \Real{3}$.
\begin{align}
\bar{y} = X_i\inv \bar{l} + \tilde{C} \mathring{\delta}
\label{eq:landmark_measurement_def}
\end{align}
where $ C \in \Real{3 \times 3}$ is determined by the sensor properties. 

A similar sensor on each robot also measures the relative translation to other robots in the network. A robot, $i$, senses and identifies a known marker point, $m_j$, affixed to another robot, $j$. The measurement, $z_{ij} \in \Real{3}$, is corrupted by zero-mean sensor noise, $\eta \in \Real{3}$.
\begin{align}
\bar{z}_{ij} = X_i\inv X_j \bar{m}_j  + \tilde{D} \mathring{\eta}
\label{eq:robot_measurement_def}
\end{align}
where $ D \in \Real{3 \times 3}$ is determined by the sensor properties. The marker point, $m_j$, is known and is defined with respect to the body-fixed frame of robot $j$.

Landmark and robot measurements are not necessarily available at all times or to all robots. Measurements may be intermittent and a robot may only be able to observe a subset of $L$ and $N$ at any given time.

\subsection{Global State Formulation}
We introduce the global state variable, $\bm{X}$, which comprises of the states of all robots in the network.
\begin{align}
\bm{X} &:= \left( X_1,\ldots, X_n \right) \in \SE(3)^n
\intertext{We then have}
\dot{\bm{X}}
&:= \bm{X}  \odot \left(\Omega_1, \ldots, \Omega_n\right), 
& \bm{X}(0) &= \bm{X}_0,
\label{eq:global_kinematics}
\end{align}
and denote
\begin{align}
\bm{u} &:= \left(u_1, \ldots, u_n\right),
&
\bm{\epsilon} &:= \left(\epsilon_1, \ldots, \epsilon_n\right).
\end{align}

\subsection{Cost Functional}
As discussed in Section \ref{sec:measurements}, each robot can receive information from three different sensors to provide measurements of velocity, positions of landmarks and positions of other robots. We follow the approach taken by \cite{zamaniDiscreteUpdatePose2019} to define the problem in terms of a continuous-time propagation step that uses the velocity measurements, and a discrete time update step, using either the landmark or robot measurements.

Following Mortensen's formulation of the deterministic minimum energy problem \citep{mortensenMaximumlikelihoodRecursiveNonlinear1968}, we introduce the following continuous-time cost functional, $J_t$.
\begin{gather}
J_t(\bm{X}, \bm{\epsilon}) := \frac{1}{2} d_{P_0}^2\left(\bm{X}(0), \hat{\bm{X}}_0\right) + \frac{1}{2} \sum_{i\in N} \int_{0}^{t}  \norm{\epsilon_i}^2 d\tau
\end{gather}
where $P_0$ is a positive definite matrix which weights the initial estimate.
We assume that, relative to the exteroceptive measurements, the velocity measurements are available at a high enough frequency that they can be regarded as a continuous signal.

We can now formally define the minimum energy filtering problem: Given a sequence of velocity measurements, $\bm{u}[0,t]$, find an estimate, $\hat{\bm{X}}(t) \in \SE(3)^n$, of the state of the system, $\bm{X}(t)$, that minimises the cost functional $J_t(\hat{\bm{X}}, \bm{\epsilon})$ and is consistent with the kinematics described in~\eqref{eq:global_kinematics}. The estimate must also be formulated as a recursive equation, dependent only on the measurements and the state estimate at the current time.

Minimising $J_t$ is performed in two steps -- firstly by minimising over $\bm{\epsilon}$, and then minimising over a point $\bm{X}$ on the trajectory. We introduce the value function, $V$, to represent the first step in this process.
\begin{align}
&V(\bm{X}, t) := \min_{\bm{\epsilon}[0,t]} J_t(\bm{X}, \bm{\epsilon})
\\
&V(\bm{X}(0), 0) = \frac{1}{2} d_{P_0}^2\left(\bm{X}(0) - \hat{\bm{X}}_0\right)
\end{align}

The optimal state estimate is then given by
\begin{gather}
\hat{\bm{X}}(t) := \argmin_{\bm{X}} V(\bm{X}, t).
\label{eq:def_x_hat}
\end{gather}
We now consider the exteroceptive landmark and robot measurements. As in \cite{zamaniDiscreteUpdatePose2019}, we introduce a discrete-update value function, $V^+$, for landmark measurements. Additionally, we introduce a second value function, $V^{\doubleplus}$, for measurements of other robots.
\begin{align}
V^+(\bm{X}, t) &:= V(\bm{X}, t) + \frac{1}{2} \norm{X_i \bar{y}_i - \bar{l} }^2_{P_{\bar{y}}\inv}
\label{eq:def_V_p}
\\
V^{\doubleplus}(\bm{X}, t) &:= V(\bm{X}, t) + \frac{1}{2} \norm{X_i \bar{z}_{ij} - X_j\bar{m}_j }^2_{P_{\bar{z}}\inv}
\\
\mathrlap{
P_{\bar{y}} := \tilde{C} \tilde{C}^\top,  \quad P_{\bar{z}} := \tilde{D} \tilde{D}^\top
}
\end{align}
The optimal minimum-energy state estimate is given by $\hat{\bm{X}}^+(t)$ or $\hat{\bm{X}}^{\doubleplus}(t)$, respectively
\begin{align}
\hat{\bm{X}}^+(t) &:= \argmin_{\bm{X}} V^+(\bm{X}, t)
\label{eq:def_x_hat_p}
\\
\hat{\bm{X}}^{\doubleplus}(t) &:= \argmin_{\bm{X}} V^{\doubleplus}(\bm{X}, t)
\end{align}
While these equations have been formulated for a single measurement, they are applied to each landmark or robot measurement at the time they are received.

%% file: sections/4_results.tex
\section{Results}
\label{sec:results}
In this section, we derive the filter equations for the centralised state estimation problem and then show how they can be decoupled and distributed among the robots in the network.

\subsection{Central GAME Filter Formulation}
\label{sec:central_game}
The following lemma is a simple consequence of the relevant definitions.
\begin{lem}
	\label{lem:hess}
	Given any two tangent directions $\bm{X \Gamma}, ~ \bm{X \Psi} \in T_{\bm{X}}\SE(3)^n$, the Hessian of the value function, 
	acting as a symmetric mapping with respect to the inner product	is equivalently represented with a positive definite matrix, $P \in \Real{6n\times6n}$, operating on vectors $\bm{\Gamma}^{\curlyvee}, \bm{\Psi}^{\curlyvee} \in \Real{6n}$.
	\begin{gather}
	\left\langle P \bm{\Psi}^\curlyvee, \bm{\Gamma}^\curlyvee \right\rangle := \left\langle \Hess_{\bm{X}} V(\bm{X}, t) \circ \bm{X\Psi}, \bm{X\Gamma} \right\rangle _{|\bm{X} = \hat{\bm{X}}(t)}
	\label{eq:def_P}
	\end{gather}
\end{lem}

\subsubsection{Propagation of Velocity Measurements.}
Following the methodology in Theorem 1 and Theorem 2 of \cite{zamaniDiscreteUpdatePose2019} results in the following filter state propagation equations for the centralised system.
\begin{gather}
\dot{\hat{\bm{X}}}(t) = \hat{\bm{X}}(t) \odot \bm{u}(t)
\label{eq:central_x_dot}
\end{gather}
\begin{align}
\dot{P}(t) &= -P\bm{BB}^\top P + \projsym{P\bm{U}}, & P(0) &= P_0
\label{eq:central_P_dot}
\intertext{where}
U &:= \begin{bmatrix} (u_\omega)_\times & 0 \\ (u_v)_\times & (u_\omega)_\times \end{bmatrix},
\\
\bm{U} &:= \blkdiag(U_1, U_2, \ldots, U_n),
\\
\bm{B} &:= \blkdiag(B_1, B_2, \ldots, B_n).
\end{align}

We note here that $\bm{B}$ is block diagonal, indicating our assumption that velocity measurements on-board one robot are independent of all other robots.

\subsubsection{Landmark Measurement Update}
\begin{thm}
	\label{thm:central_landmark_update}
	Consider a single relative position measurement of a landmark, $y$, as defined in \eqref{eq:landmark_measurement_def}. The approximate minimum-energy recursive solution to the estimate of the state $\bm{X}$, as defined in \eqref{eq:def_x_hat_p} is
\begin{gather}
\hat{\bm{X}}^+ = \hat{\bm{X}} \odot \bm{\Theta}
\label{eq:x_hat_p_derivation}\\
\intertext{where}
\bm{\Theta} = \exp \left(\left( -(P^+)\inv \left(\hat{\bm{X}}\inv  \nabla_{\bm{X}} V^+(\hat{\bm{X}}(t), t ) \right) ^\curlyvee \right) ^\curlywedge \right)
\end{gather}
\begin{align}
\nabla_{X_i} V^+(\hat{\bm{X}}(t), t) &= \hat{X}_i \proj{\hat{X}_i^\top P_{\bar{y}}\inv (\hat{X}_i \bar{y} - \bar{l}) \bar{y}^\top \tilde{(2 \bm{I}_3)}
 }
\label{eq:grad_x_i_v_p}
\\
\nabla_{X_j} V^+(\hat{\bm{X}}(t), t) &= 0 \quad \forall j \neq i.
\end{align}
$P^{+}$ is the matrix equivalent to $\Hess_{\bm{X}}V^{+}(\hat{\bm{X}}(t), t)$, as defined in Lemma \ref{lem:hess}, and can be calculated as
\begin{align}
P^+ &= P + Q
\\
\begin{split}
Q_{ii} &= \projsym{F(\bar{y})^\top G(\hat{X}_i^\top P_{\bar{y}}\inv (\hat{X}_i \bar{y} - \bar{l} ))} \\ &\qquad+ F(\bar{y})^\top \hat{X}_i^\top P_{\bar{y}}\inv \hat{X}_i F(\bar{y})
\end{split}
\\
Q_{k} &= \bm{0} \quad \forall k \neq (i,i).
\end{align}
 Here, $Q \in \Real{6n \times 6n}$  is indexed in blocks of $6\times6$ elements, so that $Q_{ii}$ refers to the $i$th $6\times6$ block matrix on the diagonal of $Q$. Recall the definitions of $F$ and $G$ from \eqref{eq:def_F} and \eqref{eq:def_G}, respectively.

\textit{Proof:~~} 
We first perform a Taylor expansion of $V^+(\bm{X},t)$  to second order around the point $\bm{X}=\hat{\bm{X}}(t)$ along the geodesic $\bm{\Psi} = \log(\hat{\bm{X}}\inv\odot\bm{X})$. This yields an approximate solution as the value function is not guaranteed to be of second order. Ignoring the higher order terms,
\begin{gather}
\begin{multlined}
V^+(\bm{X}, t) = V^+(\hat{\bm{X}}, t) + \langle \nabla_{\bm{X}} V^+(\hat{\bm{X}}, t), \hat{\bm{X}} \bm{\Psi} \rangle\\
+ \frac{1}{2}\langle \Hess_{\bm{X}} V^+(\hat{\bm{X}}, t) \circ \bm{\hat{X}}  \bm{\Psi}, \bm{\hat{X}} \bm{\Psi} \rangle 
\end{multlined}
\label{eq:taylor_expansion}
\end{gather}

As a consequence of \eqref{eq:def_x_hat_p}, we have
\begin{gather}
\{\mathcal{D}_{\bm{X}} V^+(\bm{X}, t) \circ \bm{X} \bm{\Gamma} \} _{\bm{X} = \hat{\bm{X}}^+(t)} = 0
\label{eq:V_p_derivative}
 \end{gather}
which we can substitute in \eqref{eq:taylor_expansion}. Combined with the consequence from \eqref{eq:def_x_hat} that $\nabla_{\bm{X}}  V(\hat{\bm{X}}, t) = 0$, it follows that
\begin{gather}
\begin{multlined}
0=\Big\{
\left\langle \nabla_{\bm{X}} V^+ (\hat{\bm{X}}(t), t),  \mathcal{D}_{\bm{X}} (\hat{\bm{X}}  \bm{\Psi}) \circ (\hat{\bm{X}} \bm{\Gamma}) \right\rangle
\\
+ \Big\langle \Hess_{\bm{X}} V^+(\hat{\bm{X}}(t), t) \circ \hat{\bm{X}}  \bm{\Psi}, \\ \mathcal{D}_{\bm{X}} (\hat{\bm{X}}  \bm{\Psi}) \circ (\hat{\bm{X}} \bm{\Gamma}) \Big\rangle
\Big\}_{\bm{X} = \hat{\bm{X}}^+(t)}
\end{multlined}
\end{gather}
Rearranging to solve for $\hat{\bm{X}}^+(t)$, together with \eqref{eq:def_P}, results in \eqref{eq:x_hat_p_derivation}. Equation \eqref{eq:grad_x_i_v_p} is then derived by evaluating
\begin{align}
\langle \nabla_{X_i} V^+(\hat{\bm{X}}(t), t), \hat{X}_i \Gamma \rangle
\end{align}
using \eqref{eq:def_derivative}, \eqref{eq:def_V_p}, and \eqref{eq:V_p_derivative}. We then evaluate the derivative and reformulate in terms of \eqref{eq:def_trace_norm} to solve for $\nabla_{X_i} V^+(\hat{\bm{X}}(t), t)$. We calculate $P^+$ by observing that
\begin{align}
\begin{split}
\Hess_{\bm{X}}V^+(\hat{\bm{X}}(t), t) &= \Hess_{\bm{X}}V(\hat{\bm{X}}(t), t)\\ &\quad + \Hess_{\bm{X}} \left( \frac{1}{2} \norm{X_i \bar{y}_i - \bar{l} }^2_{P_{\bar{y}}\inv} \right).
\end{split}
\end{align}
\begin{flushright}
\qed
\end{flushright}
\end{thm}

\subsubsection{Robot Measurement Update}
\begin{thm}
	In the same manner as Theorem \ref{thm:central_landmark_update}, the approximate minimum-energy recursive estimate for the state $\bm{X}$ after a robot measurement, $z_{ij}$, can be calculated as
\begin{gather}
\hat{\bm{X}}^{\doubleplus} =  \hat{\bm{X}} \odot \bm{\Xi}
\label{eq:central_r2r_X_update}
\intertext{where}
\bm{\Xi} =  \exp \left(\left( -(P^{\doubleplus})\inv \left(\hat{\bm{X}}\inv \odot \nabla_{\bm{X}} V^{\doubleplus}(\hat{\bm{X}}(t), t ) \right) ^\curlyvee \right) ^\curlywedge \right)
\label{eq:central_r2r_X_update_delta}
\end{gather}
\begin{align}
\nabla_{X_i} V^{\doubleplus}(\hat{\bm{X}}(t), t) &= \hat{X}_i \proj{\hat{X}_i^\top P_{\bar{z}}\inv \left(\hat{X}_i \bar{z} - \hat{X}_j \bar{m}\right) \bar{z}^\top 
	\tilde{(2 \bm{I}_3)}
}
\label{eq:grad_v_pp_i}
\\
\nabla_{X_j} V^{\doubleplus}(\hat{\bm{X}}(t), t) &= \hat{X}_j \proj{\hat{X}_j^\top P_{\bar{z}}\inv \left(\hat{X}_i \bar{z} - \hat{X}_j \bar{m}\right) \bar{m}^\top 
	\tilde{(2 \bm{I}_3)}
}
\label{eq:grad_v_pp_j}
\\
\nabla_{X_k} V^{\doubleplus}(\hat{\bm{X}}(t), t) &= 0 \quad \forall k \notin \{i,j\}
\end{align}
$P^{\doubleplus}$ is the matrix equivalent to $\Hess_{\bm{X}}V^{\doubleplus}(\hat{\bm{X}}(t), t)$, as defined in Lemma \ref{lem:hess}, and can be calculated as
\begin{align}
P^{\doubleplus} &= P + W
\label{eq:central_r2r_p_update}
\\
\begin{split}
W_{ii} &= \projsym{F(\bar{z})^\top G(\hat{X}_i^\top P_{\bar{z}}\inv (\hat{X}_i \bar{z} - \hat{X}_j\bar{m} ))}
\\
&\qquad+ F(\bar{z})^\top \hat{X}_i^\top P_{\bar{z}}\inv \hat{X}_i F(\bar{z})
\end{split}
\label{eq:W_ii}
\\
W_{ij} &= F(\bar{z})^\top \hat{X}_i^\top P_{\bar{z}}\inv \hat{X}_j F(\bar{m})
,\qquad
W_{ji} = W_{ij}^\top
\\
\begin{split}
W_{jj} &= \projsym{F(\bar{m})^\top G(\hat{X}_j^\top P_{\bar{z}}\inv (\hat{X}_j \bar{m} - \hat{X}_i\bar{z} ))}
\\
 &\qquad+ F(\bar{z})^\top \hat{X}_j^\top P_{\bar{z}}\inv \hat{X}_j F(\bar{m})
\end{split}
\label{eq:W_jj}
\\
W_{k} &= \bm{0} \quad \forall k \notin \{(i,i), (i,j), (j, i), (j,j)\}
\label{eq:W_k}.
\end{align}
Here, $W \in \Real{6n \times 6n}$ is indexed in the same way as $Q$ from the previous section.

\textit{Proof:}~~ The proof follows along the same lines as the proof of Theorem \ref{thm:central_landmark_update}.
\end{thm}
\subsection{Decoupled Central GAME Filter Formulation}
\label{sec:decoupled_game}
Given the set of equations that define the propagation and update steps for the centralised GAME filter, we now attempt to decouple the equations so that the calculations can be distributed among the robots in the network. We will find that the decoupling of the filter equations is easier when working with the inverse of the Hessian, $\Sigma := P \inv$.

In the following formulation each robot, $i$, tracks its own state estimate, $\hat{X}_i$, and an $n\times 6$ sub-matrix of $\Sigma$, $\Sigma_{ki}$.

\subsubsection{Propagation Step}

We note that the state propagation equation \eqref{eq:central_x_dot} is trivial to decouple.
\begin{gather}
\dot{\hat{X}}_i(t) = \hat{X}_i(t)u_i(t) \quad \forall i \in N
\end{gather}
To decouple the calculation for $P$, we reformulate \eqref{eq:central_P_dot} in terms of $\Sigma$, which allows us to separate $\dot{\Sigma}$ into components.
\begin{align}
\dot{\Sigma} &= \bm{BB}^\top - \projsym{\bm{U}\Sigma}
\\
\dot{\Sigma}_{ii} &= B_i B_i^\top - \projsym{U_i \Sigma_{ii}}
\\
\dot{\Sigma}_{ij} &= -\frac{1}{2} \left( U_i \Sigma_{ij} + \Sigma_{ij} U_j^\top \right)
\label{eq:decoupled_sigma_ij_dot}
\end{align}
Here, we use the same $6\times 6$ block indexing as in previous sections. We observe that the diagonal sub-matrices, $\dot{\Sigma}_{ii}$, only depend on data local to robot $i$, while the off-diagonal sub-matrices, $\dot{\Sigma}_{ij}$, depend on data local to both robot $i$ and $j$.
Given that \eqref{eq:decoupled_sigma_ij_dot} is a homogeneous linear ODE, we can find an explicit solution if we assume that $U_i$ and $U_j$ are constant.
\begin{equation}
\Sigma_{ij}(t) = \exp\left(-\frac{t}{2}U_i\right) \Sigma_{ij}(0) \exp\left(-\frac{t}{2} U_j^\top\right)
\end{equation}
In reality, the velocity measurements are received from a sensor which updates at a fixed time interval, which means that $U_i$ and $U_j$ do remain constant for a time $\Delta t$, which represents the time between two successive measurements.
Thus, we can recursively evaluate $\Sigma_{ij}$ at a time $t_n$, after the $n$-th measurement is recorded by
\begin{gather}
\mbox{$\Sigma_{ij}(t_n) =  \exp\left(-\frac{\Delta t}{2}U_i(t_n)\right) \Sigma_{ij}(t_{n-1}) \exp\left(-\frac{\Delta t}{2} U_j(t_n)^\top\right)$}\\
= K_i(t_n) \Sigma_{ij}(0) K_j(t_n)^\top
\\ K_i(t_n) = \prod_{k=n}^1 \exp\left(-\frac{\Delta t}{2}U_i(t_k)\right)
\end{gather}

Based on this formulation, we observe that $K_i$ can be computed independently by robot $i$ and similarly $K_j$ can be computed by robot $j$.
Robot $i$ can calculate $\Sigma_{ji}(t_n)$ by receiving a message from robot $j$ that contains $K_j(t_n)$.

In this way, we show a parallel result to \cite{roumeliotisDistributedMultirobotLocalization2002} whereby robots can propagate their state independently and only need to share information at a time where exteroceptive measurements are taken. 

\subsubsection{Robot Measurement Update}
In the robot measurement update step, we are required to decouple \eqref{eq:central_r2r_X_update} and \eqref{eq:central_r2r_p_update}. This would be a straightforward task for \eqref{eq:central_r2r_p_update} if $P$ was known, however given that the propagation step has been computed in terms of $\Sigma$, this would require a full matrix-inversion of $\Sigma$, which is only possible in a centralised system --- not in our system where each robot is only tracking a sub-matrix. Thus, the update step must also be reformulated and then decoupled in terms of $\Sigma$.
\begin{gather}
\Sigma^{\doubleplus} = \left(\bm{I}_{6n} + \Sigma W\right)\inv \Sigma
\label{eq:sigma_pp}
\end{gather}

Recall the definition of $W$ from \eqref{eq:W_ii} through \eqref{eq:W_k}. We observe that, because of the sparsity of $W$, the only elements of $\Sigma$ that need to be known in order to compute $(\bm{I}_{6n} + \Sigma W)\inv$ are $\Sigma_{ki}$ and $\Sigma_{kj}$, $k \in N$. This corresponds to the elements of $\Sigma$ that are being tracked by robot $i$ and $j$, respectively, and means that the inverse can be computed locally between robot $i$ and $j$. Once calculated, this term can then be shared with all other robots in the network to calculate the value for $\Sigma^{\doubleplus}$.


Similarly, \eqref{eq:grad_v_pp_i} and \eqref{eq:grad_v_pp_j} can be computed locally between robots $i$ and $j$, which also allows \eqref{eq:central_r2r_X_update_delta} to be computed locally. Each component of $\bm{\Xi}$ can then be communicated to the relevant robot such that \eqref{eq:central_r2r_X_update} can be computed locally to each robot.
\begin{gather}
\hat{X}_i^{\doubleplus} = \hat{X}_i \Xi_i \quad \forall i \in N
\end{gather}

One of the issues with the current formulation is that \eqref{eq:sigma_pp} requires inverting a $6n \times 6n$ matrix. However, we observe that $\rank(W) \leqslant 12$ and we perform a singular value decomposition on $\Sigma W$ as follows.
\begin{gather}
TSV^\top = \Sigma W
\label{eq:svd}
\\
TT^\top = VV^\top = \bm{I}_{6n}, \quad S \in \mathbb{D}^{12}, \quad T,V \in \Real{6n \times 12}
\end{gather}

We can then apply the matrix inversion lemma (Woodbury matrix identity) which reduces the size of the matrix that is inverted from $6n \times 6n$ to a maximum of $12\times 12$, depending on the actual rank of $W$. $S$ is diagonal and can be trivially inverted.
\begin{gather}
\Sigma^{\doubleplus} = \left(\bm{I}_{6n} - T (S\inv + V^\top T)\inv V^\top \right) \Sigma
\end{gather}

This also reduces the size of messages that need to be communicated, as instead of sending a $6n \times 6n$ matrix, $T$, $S$, and $V$ can be sent individually, which is only $6n \times 24 + 12$ elements.

%
%
%

\subsubsection{Landmark Measurement Update}
The decoupling of the landmark measurement update equations follows in a similar way to the previous section, giving
\begin{align}
\Sigma^+ &= (\bm{I}_{6n} + \Sigma Q)\inv \Sigma.
\end{align}

Calculating $\Sigma Q$ only requires $\Sigma_{ki},~ k\in N$ to be known. Thus robot $i$ can perform the matrix inversion locally, and then communicate the required information for all other robots to update their state and respective components of $\Sigma^+$. Similarly, $\bm{\Theta}$ can be calculated by robot $i$ and distributed to each robot to perform the update of the state estimate locally by
\begin{align}
\hat{X}_i^+ &= \hat{X}_i \Theta_i \quad \forall i \in N.
\end{align}

If we perform an SVD of $\Sigma Q$ in a similar way to \eqref{eq:svd}, we note that $\rank(Q) \leq 6$ and thus the resulting decomposition produces matrices maximally of size $6n \times 6$ and a maximum total message size of $6n \times 12 + 6$.

%% file: sections/5_simulations.tex
\section{Simulations}
\label{sec:simulations}

We demonstrate the performance of the resulting filter in two Python simulations.\footnote{Code is available at \texttt{jackhenderson.com.au}} The first considers the case where the robots' poses are constrained to a \mbox{2-D} plane, such as in the case of a network of ground based robots, while the second scenario considers the more general case of 3-D trajectories.


\subsection{2-D Case}
We consider a network of $n=4$ robots moving along circular trajectories within an approximately $20\text{m} \times 20\text{m}$ area.
There are $r=4$ landmarks in the environment and each robot is only able to take measurements of one distinct landmark at a rate of 10 Hz.
Velocity measurements are available to each robot at a rate of 100 Hz. Robots can observe only a single other distinct robot at a rate of 5 Hz but can communicate freely to all. To be specific, Robot 1 can observe Robot 2, R2 can observe R3, R3 can observe R4, and R4 can observe R1.
The sensor properties are defined as
\begin{align}
B &= 0.05 \bm{I}_6,& ~C &= 0.5 \bm{I}_3,& ~ D &= 0.5\bm{I}_3,
\label{eq:sim_noise_1}
\\
\epsilon & \sim \mathcal{N}(\bm{0},\bm{I}_6), & \delta  & \sim \mathcal{N}(\bm{0},\bm{I}_3), & \eta  & \sim \mathcal{N}(\bm{0},\bm{I}_3).
\label{eq:sim_noise_2}
\end{align}
Note, the sensor errors are constrained appropriately in the 2-D case.

\begin{figure}
	\includegraphics[width=\linewidth]{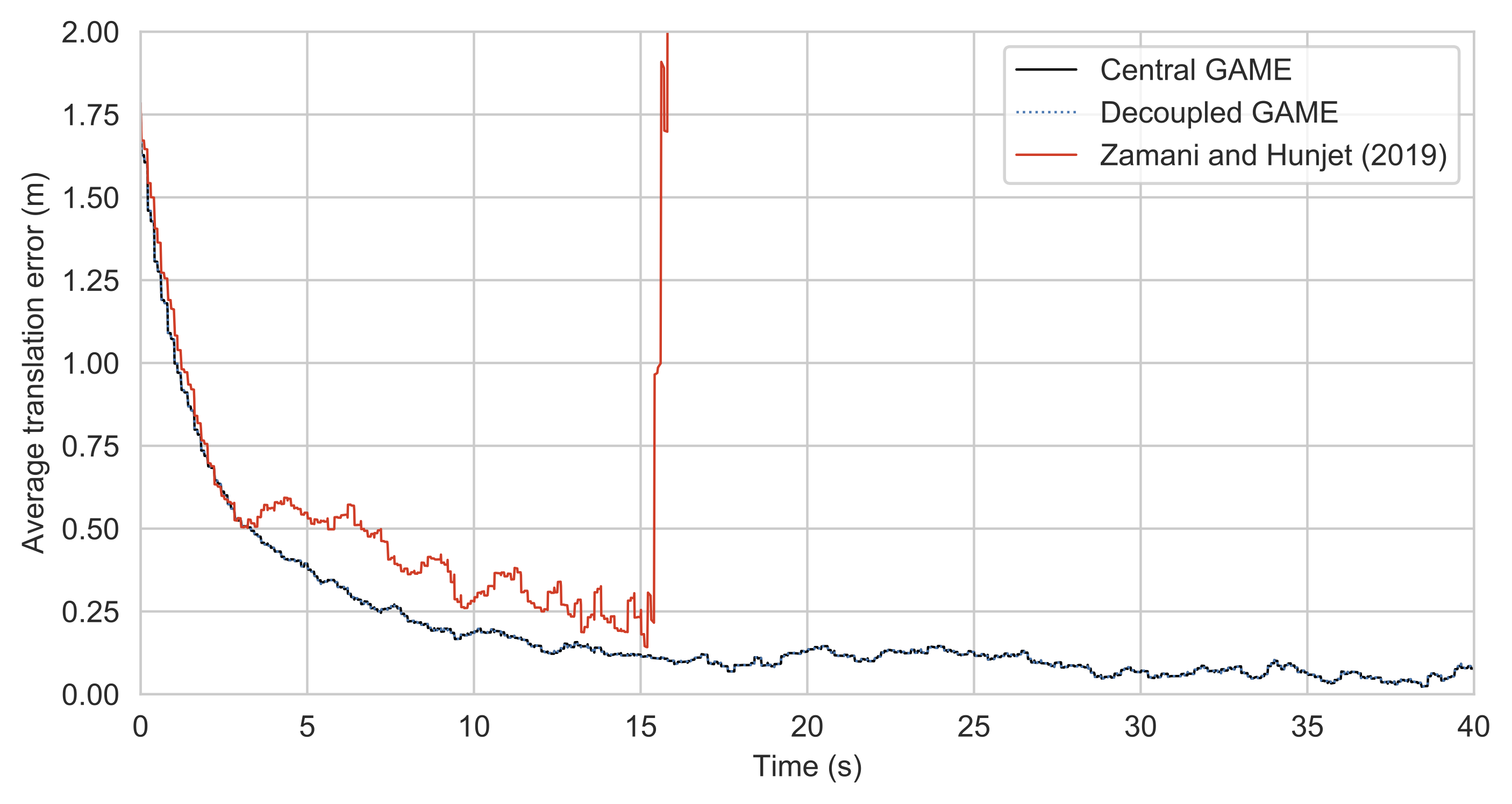}
	\caption{Average translation error across all robots for 2-D Scenario. Best viewed in colour.}
	\label{fig:translation_error}
\end{figure}

We implement three filters, the centralised GAME filter described in Section \ref{sec:central_game}, the decoupled GAME filter described in Section \ref{sec:decoupled_game}, and the collaborative GAME filter proposed by \cite{zamaniCollaborativePoseFiltering2019}, using the Covariance Intersection estimation method.
The average translation error over all robots is shown in Figure \ref{fig:translation_error}.

We can observe that the centralised GAME filter and the decoupled GAME filter provide identical state estimates, demonstrating that there is no loss of information when the centralised filter is decoupled. Our filter is able to accurately localise the network of robots from an initial average translation error of 1.8m down to a long term average of 0.08m.  This simulation also highlights a weakness in the filter from \cite{zamaniCollaborativePoseFiltering2019} which stems from the asymmetry of the robot observations. As it does not share the information gained from measurements to other robots, the filter is not able to accurately localise and it diverges after approximately 15 seconds.

\subsection{3-D Case}
\begin{figure}
	\includegraphics[width=\linewidth]{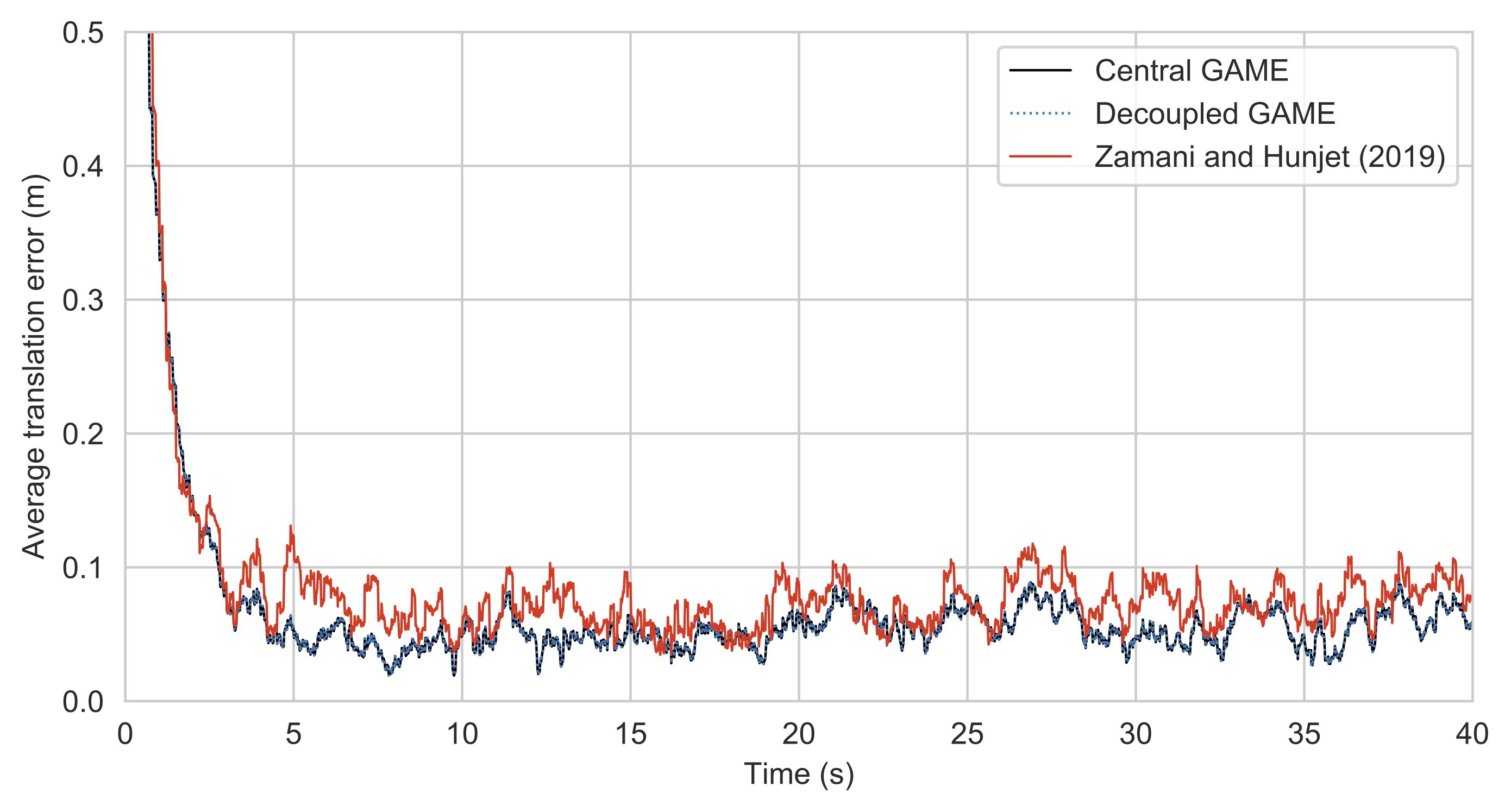}
	\caption{Average translation error across all robots for 3-D Scenario. Best viewed in colour.}
	\label{fig:translation_error_3D}
\end{figure}

We present a different scenario to highlight the difference in filter performance even when measurements are readily available to all robots. We again consider a network of $n=4$ robots with $r=4$ fixed landmarks in the environment. Velocity measurements are available to each robot at a rate of 100 Hz. In contrast to the previous scenario, each robot can observe all 4 landmarks at a rate of 10 Hz, and can observe all other robots at a rate of 10 Hz. Robots move with continuously changing random velocities in an approximately $20\text{m} \times 20 \text{m} \times 20 \text{m}$ volume. The sensor properties are the same as defined in \eqref{eq:sim_noise_1} and \eqref{eq:sim_noise_2}. The average translation error of the three different filters is shown in Figure \ref{fig:translation_error_3D}. The covariance intersection method in the \cite{zamaniCollaborativePoseFiltering2019} filter was coarsely tuned to a value of $\omega=0.03$.

Given the abundance of landmark measurements, both filters localise rapidly from the original initialisation error of 1.8m. However,  the \cite{zamaniCollaborativePoseFiltering2019} filter converges to a long-term average error of 0.074m, compared to 0.053m for our filter.

%% file: sections/6_conclusions.tex
\section{Conclusion}
\label{sec:conclusion}

In this paper, we have shown how minimum energy filtering can be applied to the collaborative localisation problem. We demonstrate how our centralised filter can be decoupled and distributed among the robots in the network. The simulations presented verify that no information is lost when distributing the filters and demonstrates superior performance compared with previous work.

Planned future work includes a comparison to similar EKF filters and a demonstration on real hardware. While the all-to-all communication requirements for this algorithm may be too restrictive for many scenarios, we can use the algorithm developed here to guide further research. In the same way that \cite{roumeliotisDistributedMultirobotLocalization2002} has been used as the benchmark for further improvements to EKF localisation algorithms, we can use our filter as the benchmark to quantify the reduction in performance that comes with reducing communication constraints for minimum energy filters.

\color{black}

%% file: main.bbl
\begin{thebibliography}{9}
\providecommand{\natexlab}[1]{#1}
\providecommand{\url}[1]{\texttt{#1}}
\providecommand{\urlprefix}{URL }
\expandafter\ifx\csname urlstyle\endcsname\relax
  \providecommand{\doi}[1]{doi:\discretionary{}{}{}#1}\else
  \providecommand{\doi}{doi:\discretionary{}{}{}\begingroup
  \urlstyle{rm}\Url}\fi

\bibitem[{{Carrillo-Arce} et~al.(2013){Carrillo-Arce}, Nerurkar, Gordillo, and
  Roumeliotis}]{carrillo-arceDecentralizedMultirobotCooperative2013}
{Carrillo-Arce}, L.C., Nerurkar, E.D., Gordillo, J.L., and Roumeliotis, S.I.
  (2013).
\newblock Decentralized multi-robot cooperative localization using covariance
  intersection.
\newblock In \emph{2013 {{IEEE}}/{{RSJ International Conference}} on
  {{Intelligent Robots}} and {{Systems}}}, 1412--1417.

\bibitem[{Howard et~al.(2003)Howard, Mataric, and
  Sukhatme}]{howardPuttingTeamEgocentric2003}
Howard, A., Mataric, M., and Sukhatme, G. (2003).
\newblock Putting the `{{I}}' in `team': An ego-centric approach to cooperative
  localization.
\newblock In \emph{2003 {{IEEE International Conference}} on {{Robotics}} and
  {{Automation}}}, 868--874.

\bibitem[{Luft et~al.(2018)Luft, Schubert, Roumeliotis, and
  Burgard}]{luftRecursiveDecentralizedLocalization2018}
Luft, L., Schubert, T., Roumeliotis, S.I., and Burgard, W. (2018).
\newblock Recursive decentralized localization for multi-robot systems with
  asynchronous pairwise communication.
\newblock \emph{The International Journal of Robotics Research}, 37(10),
  1152--1167.

\bibitem[{Markley(2003)}]{markleyAttitudeErrorRepresentations2003}
Markley, F.L. (2003).
\newblock Attitude {{Error Representations}} for {{Kalman Filtering}}.
\newblock \emph{Journal of Guidance, Control, and Dynamics}, 26(2), 311--317.

\bibitem[{Mortensen(1968)}]{mortensenMaximumlikelihoodRecursiveNonlinear1968}
Mortensen, R.E. (1968).
\newblock Maximum-likelihood recursive nonlinear filtering.
\newblock \emph{Journal of Optimization Theory and Applications}, 2(6),
  386--394.

\bibitem[{Roumeliotis and
  Bekey(2002)}]{roumeliotisDistributedMultirobotLocalization2002}
Roumeliotis, S. and Bekey, G. (2002).
\newblock Distributed multirobot localization.
\newblock \emph{IEEE Transactions on Robotics and Automation}, 18(5), 781--795.

\bibitem[{Zamani and Hunjet(2019)}]{zamaniCollaborativePoseFiltering2019}
Zamani, M. and Hunjet, R. (2019).
\newblock Collaborative {{Pose Filtering Using Relative Measurements}} and
  {{Communications}}.
\newblock In \emph{12th {{Asian Control Conference}} ({{ASCC}})}, 919--924.

\bibitem[{Zamani and Trumpf(2019)}]{zamaniDiscreteUpdatePose2019}
Zamani, M. and Trumpf, J. (2019).
\newblock Discrete update pose filter on the special {{Euclidean}} group
  {{SE}}(3).
\newblock In \emph{Proceedings of the 55th {{IEEE Conference}} on {{Decision}}
  and {{Control}} ({{CDC}})}.

\bibitem[{Zamani et~al.(2013)Zamani, Trumpf, and
  Mahony}]{zamaniMinimumEnergyFilteringAttitude2013}
Zamani, M., Trumpf, J., and Mahony, R. (2013).
\newblock Minimum-{{Energy Filtering}} for {{Attitude Estimation}}.
\newblock \emph{IEEE Transactions on Automatic Control}, 58(11), 2917--2921.

\end{thebibliography}
